\documentclass[twocolumn,showpacs,pra]{revtex4-1}

\usepackage{graphicx}
\usepackage{float, amsmath, amssymb, color}

\usepackage{times}

\newcommand{\nr}{\overline{n}_R(\tau)}
\newcommand{\jp}{J_\parallel}

\begin{document}

\title{Non-equilibrium dynamics in Bose-Hubbard ladders}

\author{Wladimir Tschischik}
\author{Masudul Haque}
\author{Roderich Moessner}
\affiliation{Max-Planck-Institut f\"ur Physik komplexer Systeme, N\"othnitzer Str. 38, 01187 Dresden, Germany}

\date{\today}

\begin{abstract}

Motivated by a recent experiment on the non-equilibrium dynamics of
interacting bosons in ladder-shaped optical lattices, we report exact
calculations on the sweep dynamics of Bose-Hubbard systems in finite two-leg
ladders.  The sweep changes the energy bias between the legs linearly over a
finite time.  As in the experiment, we study the cases of [a] the bosons
initially all in the lower-energy leg (ground state sweep) and [b] the bosons
initially all in the higher-energy leg (inverse sweep).  The approach to
adiabaticity in the inverse sweep is intricate, as the transfer of bosons is
non-monotonic as a function of both sweep time and intra-leg tunnel coupling.
Our exact study provides explanations for these non-monotonicities based on
features of the full spectrum, without appealing to concepts (e.g., gapless
excitation spectrum) that are more appropriate for the thermodynamic limit.
We also demonstrate and study St\"uckelberg oscillations in the finite-size 
ladders.

\end{abstract} 

\maketitle 

\section{introduction}

The unprecedented experimental tunability of ultra-cold atomic systems in
optical potentials provides the opportunity to study non-equilibrium quantum
dynamics in previously inaccessible
regimes \cite{bloch_review,polkovnikov_review}.  Very recently, pioneering
experiments \cite{experiment,DMRG,DeMarco_PRL11} have started experimentally
exploring issues related to \emph{adiabaticity}, a fundamental concept of
quantum dynamics \cite{adiabatic_theorem}.  Of particular interest is the
quantitative characterization of deviations from ideal adiabatic behavior when
the sweep rate of a system parameter is small but not infinitesimal.
Deviation from adiabaticity in many-body quantum systems has also generated
extensive theoretical
interest \cite{finiteRateQuenches_reviews,polkovnikov_review}.  This may be
regarded as the slow-sweep counterpart of the much-studied quantum quench
(infinite sweep rate) \cite{polkovnikov_review}.

\begin{figure*}
\centering
  \includegraphics[width=.92\textwidth]{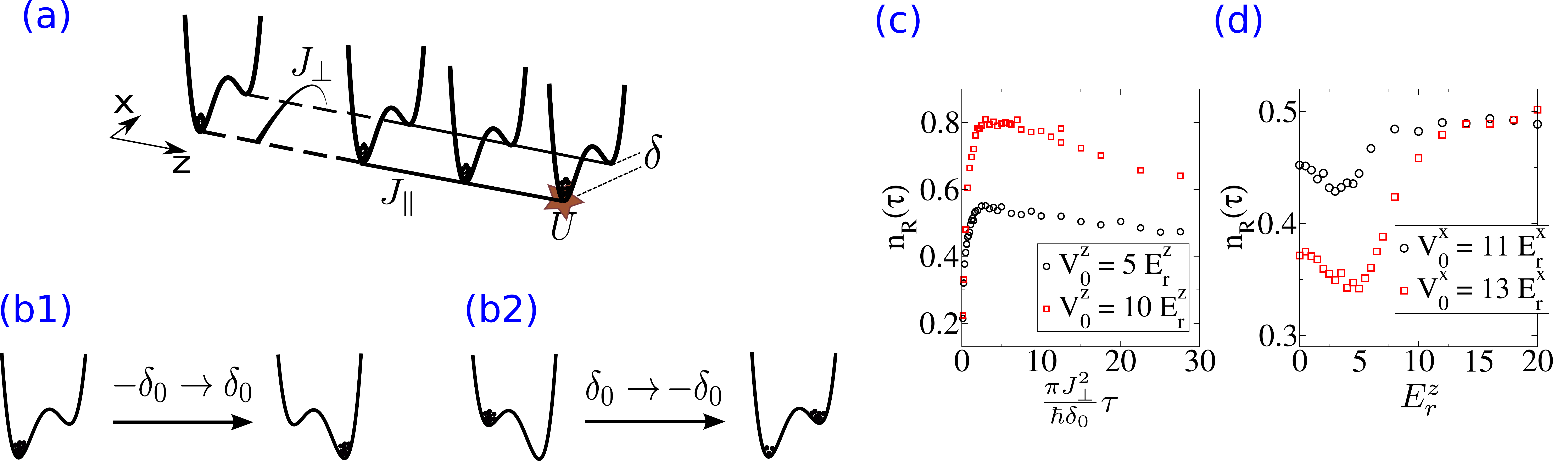}
\caption{ \label{fig:experiment_drawing}
(color online) 
(a) Each rung of our Bose-Hubbard ladder is a double-well potential (a
Bose-Hubbard dimer).  Parameters $J_\perp$, $J_\parallel$, $U$ and $\delta$
are indicated pictorially.
(b) Ground state sweep and inverse sweep. The initial state has all bosons on
the left site.  
(c,d) Representative experimental data from Ref.~\cite{experiment}, showing
two non-monotonicities of the transfer efficiency $n_R(\tau)$.
(c) Non-monotonic behavior of $n_R(\tau)$ function (``breakdown of
adiabaticity''), for different axial $z$-lattice depths.
%
%
(d) $n_R(\tau)$ at a constant value of $\tau$  
$\left[\frac{\pi{J_\perp^2}\tau}{\hbar\delta_0} = 0.53(3)\right]$ as a function of the axial
$z$-lattice depth (roughly: as a function of inverse $J_{\parallel}$), for
different inter-tube hoppings.
}
\end{figure*}

In the experiment of Ref.~\cite{experiment}, which is the motivation of the
present work, adiabaticity is explored through slow ramps of the energy bias
(relative potential energy) between two legs of a ladder.  The ladder is
formed by two coupled one-dimensional tubes or chains.  Alternatively, one can
think of the experimental setup as a ladder of dimers (double well potentials)
coupled to each other, as in Figure \ref{fig:experiment_drawing}a.  The experimental
sequence starts by loading all bosonic atoms in the left leg.  The initially
filled left leg can be lower in potential energy than the right leg
(\emph{ground state sweep}) or higher (\emph{inverse sweep}).  Subsequently
the potential energy is reversed linearly in time,
Figure \ref{fig:experiment_drawing}b.  In a truly adiabatic sweep, i.e., in
the limit of large sweep time $\tau$, it is expected that all the bosons get
transferred to the right leg.  The fraction of particles in the right leg at
the end of the sweep, $n_R(\tau)$, the so-called \emph{transfer efficiency},
characterizes adiabaticity.

For the ground state (g.s.) sweep the transfer efficiency increases with sweep
time $\tau$, as expected from the quantum adiabatic theorem.  In contrast, in
the inverse sweep case, after an initial increase $n_R(\tau)$ decreases with
$\tau$.  Following Ref.~\cite{experiment}, we call this non-monotonic behavior
a \emph{breakdown of adiabaticity}.  Of course, for extremely slow sweeps
(extremely large $\tau$), the adiabatic theorem assures us that $n_R(\tau)$
has to increase again to $n_R(\tau\rightarrow\infty)=1$.  For some $\tau$
values, $n_R(\tau)$ also has a non-monotonic dependence on the intra-leg
tunnel coupling.  The two non-monotonicities are shown through a selection of
experimental data (provided by the authors of Ref.~\cite{experiment}) in
Figures \ref{fig:experiment_drawing}(c,d).

In this work we analyze exactly the sweep dynamics in small ladder systems.
We find that already the smallest system of two bosons in a two-rung ladder
shows the two non-trivial non-monotonicities in transfer efficiency, with
respect to sweep time and with respect to intra-leg tunneling constant.
Analyzing the energy spectra of small ladder systems allows us to explain both
effects, and gives us an alternate and useful perspective on the physics
involved in sweep dynamics for the Bose-Hubbard ladder and the breakdown
phenomenon.  The sweep dynamics can be interpreted as a combination of
Landau-Zener (LZ) processes \cite{LZ} through a complex network of avoided
level crossings.  This exact study allows us to understand the main phenomena
without recourse to mean field treatments or low-energy effective
descriptions.  In addition, small-ladder systems show additional interesting
features in the dynamics (St\"uckelberg oscillations) which are averaged out
in larger systems.

In the presence of an overall trapping potential, even if the central ladders
are long, there are off-center ladders which consist of a few rungs.  Thus,
although Ref.~\cite{experiment} focuses on long ladders, the results of the
present study should directly describe some part of the experimental
measurements performed for Ref.~\cite{experiment}.

We will use the Bose-Hubbard Hamiltonian to describe the system
(Figure \ref{fig:experiment_drawing}a):
\begin{equation}
\begin{split}
 H =& - \left( J_{\perp} \sum_i b_{i,L}^\dagger b_{i,R} + J_{\parallel} \sum_{i,\sigma}  b_{i,\sigma}^\dagger b_{i+1,\sigma} \right) + \text{h.c.} \\
     &+ \frac{U}{2}\sum_{i,\sigma}\hat{n}_{i,\sigma}(\hat{n}_{i,\sigma}-1) + \frac{\delta(t)}{2}\sum_i(\hat{n}_{i,L} - \hat{n}_{i,R}),
\end{split}
\label{eq:Hamiltonian_ladder}
\end{equation}
where $b_{i,\sigma}$, $b_{i,\sigma}^\dagger$ are the bosonic operators for the
site on rung $i$ ($ i = 1 \ldots L_s$) and leg $\sigma = \text{L, R}$, and
$\hat{n}_{i,\sigma}=b_{i,\sigma}^\dagger b_{i,\sigma}$.  The length of the
ladder (number of rungs) is $L_s$.  The parameters $J_{\perp}$ and
$J_\parallel$ are the inter-leg and intra-leg tunnel couplings, respectively,
$U$ is the on-site interaction energy, and $\delta(t)$ is the time-dependent
energy bias between the two legs.  We use $J_\perp = 1$, measuring energy
[time] in units of $J_\perp$ [$\hbar/J_\perp$]. For convenience $\hbar=1$.

The initial state has all bosons on the left leg.  In our computation we
implement this by taking the initial state to be the ground state of the
Hamiltonian \eqref{eq:Hamiltonian_ladder} with $\delta=-10^{6}$.  Thus the
initial state of the time evolution is not exactly an eigenstate of the
Hamiltonian $H(\delta=\pm \delta_0)$ at $t=0$, but for relatively large
$\delta_0$ (we typically use $\delta_0$ between 20 and 40), the initial
overlap with the relevant eigenstate of $H(t=0)$ is nearly unity.

During a sweep the energy bias between left and right legs is changed linearly
in time:
\begin{equation}
 \delta(0 \leq t \leq \tau ) =  
\begin{cases} 
 \frac{2\delta_0}{\tau}t - \delta_0, & \text{ground state sweep} \\
 - \frac{2\delta_0}{\tau}t + \delta_0, & \text{inverse sweep}
\end{cases}
\label{eq:sweep_form}
\end{equation}
where $\tau$ is a sweep time.  The dynamics is characterized by the fraction
of particles on the right leg
\begin{equation}
 n_R(t) := N^{-1} \sum_i \langle \hat{n}_{i,R}(t) \rangle,
\end{equation}
where $N$ is the total number of bosons.  In our calculations, the time
evolution of the quantum state is obtained by numerical integration of the
Schr\"odinger equation and exact instantaneous eigenvalues and eigenstates of
the Hamiltonian \eqref{eq:Hamiltonian_ladder} are calculated using full
numerical diagonalization.

We start in Section \ref{sec:dimer} with the sweep dynamics of a single
Bose-Hubbard dimer, and show that the g.s.\ and inverse sweeps have different
behaviors already for the dimer.  Sections \ref{sec:ladder}
and \ref{sec:non-monotonicities} contain our main results on Bose-Hubbard
ladders.  Here we identify pertinent features of energy spectra of finite
ladders.  We present and explain non-monotonic behaviors of the transfer
efficiency $n_R(\tau)$, observed already in the smallest ladders.  In
Section \ref{sec:S_oscillations} we explain oscillations of $n_R(\tau)$
(St\"uckelberg oscillations), which appear in finite-size ladders but get
washed out in the thermodynamic limit.  In
Section \ref{sec:summary_conclusion} we summarize results and mention open
problems.  In the appendices we describe the treatment of non-interacting
Bose-Hubbard ladders and show some results for a larger system.


\section{Bose-Hubbard dimer}\label{sec:dimer}

We begin with an $L_s=1$ ladder, i.e., a single Bose-Hubbard dimer.  Spectral
properties and dynamics of the dimer has been studied by various authors
(e.g.\ \cite{Milburn,Kalosakas1, Kalosakas2,Kalosakas3, Salgueiro,Smerzi1,
Smerzi2,tejaswi, WuLiu_PRL06}).  Linear sweeps of the energy bias from
$-\delta_0$ to $0$ were studied in \cite{tejaswi} in the large $U$ limit.
Here, motivated by the expeiments \cite{experiment}, we focus on small to
intermediate $U$ and on the $\pm\delta_0\rightarrow\mp\delta_0$ sweeps shown
in Figure \ref{fig:experiment_drawing}b.
We will show that the dimer already shows marked difference between ground
state sweeps and inverse sweeps, but there is no `breakdown' phenomenon in the
inverse sweep.

\begin{figure*}[tb]
 \centering
\includegraphics[width=0.92\textwidth]{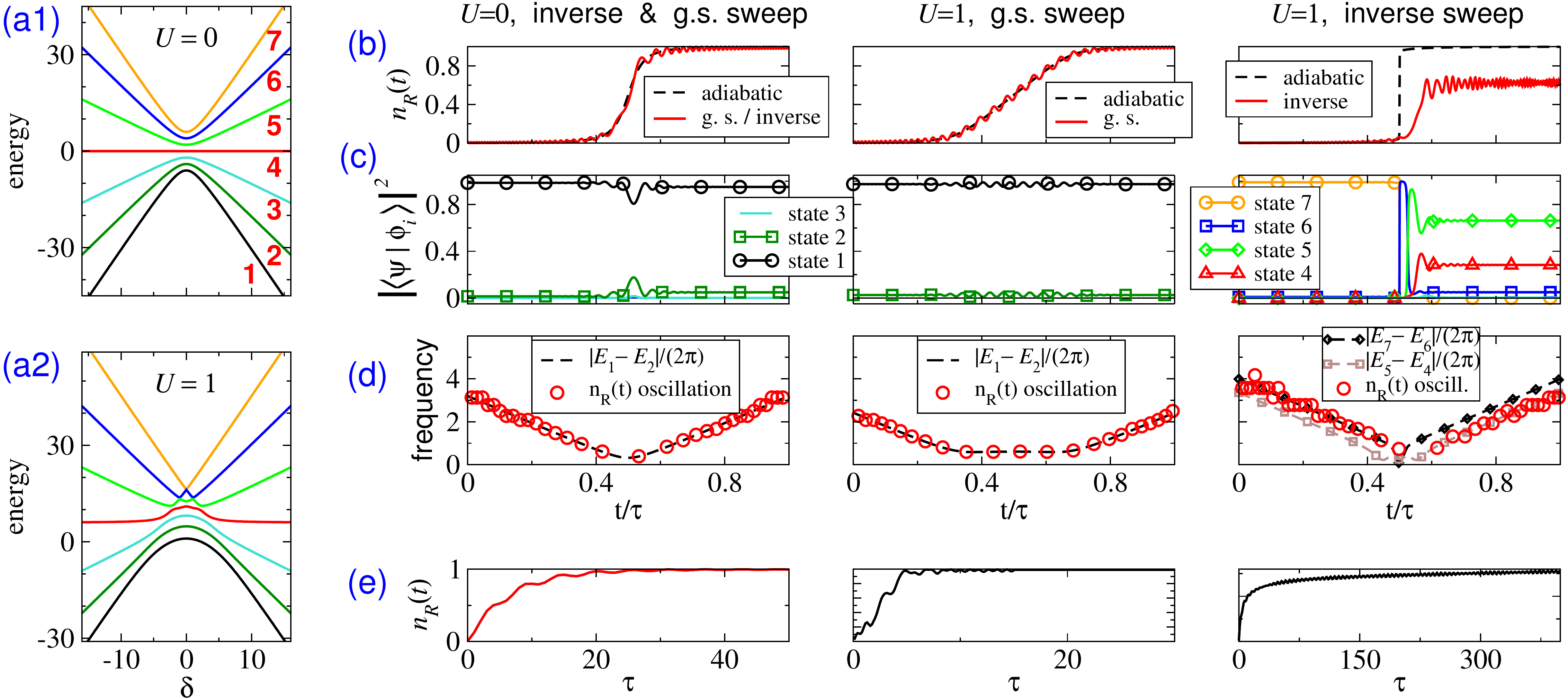}
\caption{ \label{fig:dimer_combined}
(color online) 
Spectrum and sweep dynamics of the  Bose-Hubbard dimer (single-rung ladder)
with $N=6$ bosons. 
(a) Energy spectrum for $U=0$ and $U=1$.
(b,c,d) Sweep dynamics for ($\tau$,$\delta_0$) = (40,20).  For $U=0$ (first
column), g.s.\ and inverse sweeps are identical.  State numbering is shown
for g.s\ sweeps and the inverse sweep simply has states 1,2,... replaced by
7,6,...
(b) Population fraction in right site,  $n_R(t)$. 
(c) overlap with the instantaneous eigenstates $|\phi_i\rangle$.  
(d) Frequency of oscillations of $n_R(t)$, contrasted with differences of
instantaneous energy eigenvalues.  
(e) Transfer efficiency, showing no pronounced non-monotonicity or extended
region of decrease.
}
\end{figure*}

Figure \ref{fig:dimer_combined}a shows energy spectra for the dimer.  There are $N+1$
states.  At finite $U$, the states at the top of the spectrum show a cluster
of level crossings with tiny gaps, forming a characteristic and well-known
swallow-tail shaped structure (e.g.\ \cite{experiment, odell_PRA11,
Smerzi_EPJD02, WuLiu_PRL06, GraefeKorsch_PRA07}).

Figure \ref{fig:dimer_combined}b shows the time evolution of the right site
occupancy for ground state and inverse sweeps, for relatively large sweep time
$\tau=40$.  In the non-interacting case ground state and inverse sweeps are
identical, which is expected because upper and lower parts of the spectrum are
equivalent.  The adiabaticity is quite good for $\tau=40$, as seen from
$n_R(t)$ and the overlaps with eigenstates.  

For $U > 0$ the ground state and inverse sweeps show different behavior.  For
the g.s.\ sweep, the adiabaticity is even better at the same $\tau$ than at
$U=0$, because the ground state is a bit more separated at $U=1$.  In
contrast, for the same $\tau$, during the inverse sweep several states become
populated by a sequence of Landau-Zener transitions, causing a
decrease of boson transfer to the right site.  To perform the adiabatic
transfer, a sweep has to be carried out much more slower than for a ground
state sweep.  The reason for this difference is that the energy level
structure at the top of the spectrum now contains many level crossings with
small gaps.

Oscillations in $n_R(t)$ (Figure \ref{fig:dimer_combined} panels d) are due to
occupancy of multiple eigenstates.  There are already some oscillations at
the beginning of the sweep because the initial state is an eigenstate of
$H(\delta=-10^{6})$ and not of $H(\delta=\pm \delta_0)$.  During the sweep,
eigenstates other than the lowest/highest get occupied further.  Since the
energy difference between instantaneous eigenstates (and therefore oscillation
frequencies) are changing with time during the sweep, we cannot obtain the
frequencies by a Fourier transform of $n_R(t)$; in
Figure \ref{fig:dimer_combined} panels d they are approximated from the time
difference between successive minima.

The transfer efficiency (Figure \ref{fig:dimer_combined}e) increases
monotonically (neglecting oscillations) with $\tau$, for both types of sweep.
Thus, in the dimer system inverse sweep dynamics shows no breakdown of
adiabaticity, i.e., no pronounced non-monotonicity in $n_R(\tau)$.

\section{Bose-Hubbard ladders}  \label{sec:ladder}

In this section and in section \ref{sec:non-monotonicities} we analyze the
non-equilibrium sweep dynamics of ladder systems with a few rungs.  In this
section we describe features of the energy spectrum, show what the sweeps mean
in terms of the energy spectrum, and demonstrate that the smallest possible
ladder system (two bosons in a two-rung ladder, $L_S=N=2$) already displays
the two intriguing non-monotonicities that are the main topic of this article.
(Non-monotonic dependences of $n_R(\tau)$ as a function of $\tau$ and as a
function of $\jp$, closely analogous to the two non-monotonicities found in
Ref.~\cite{experiment} and introduced in Figure \ref{fig:experiment_drawing}.)
In section \ref{sec:non-monotonicities} we provide an explanation of the two
non-monotonicities in terms of spectral features and transitions between
states of a diabatic basis.

We will use periodic boundary conditions in the leg direction.  For a ladder
containing only two dimers (rungs), this simply means doubling the intra-leg
tunneling constant $\jp$.  Systems with periodic boundary conditions are
translation invariant and the dynamics considered in this paper preserves the
total momentum along the leg direction.  Since the initial state has zero
total momentum, the dynamics is entirely confined to the zero-momentum sector.
Thus we use only the subspace of the Hilbert space spanned by the linear
combination of the position basis states of the form
$1/\sqrt{L_s} \sum_{j=0}^{L_s-1} T^j |n_{1,L},n_{1,R},\dots,n_{L_s,L},
n_{L_s,R} \rangle$, where  $T$ is the translation operator.


\subsection{Energy spectrum of Bose-Hubbard ladders}

\begin{figure}
\centering
\includegraphics[width=0.99\columnwidth]{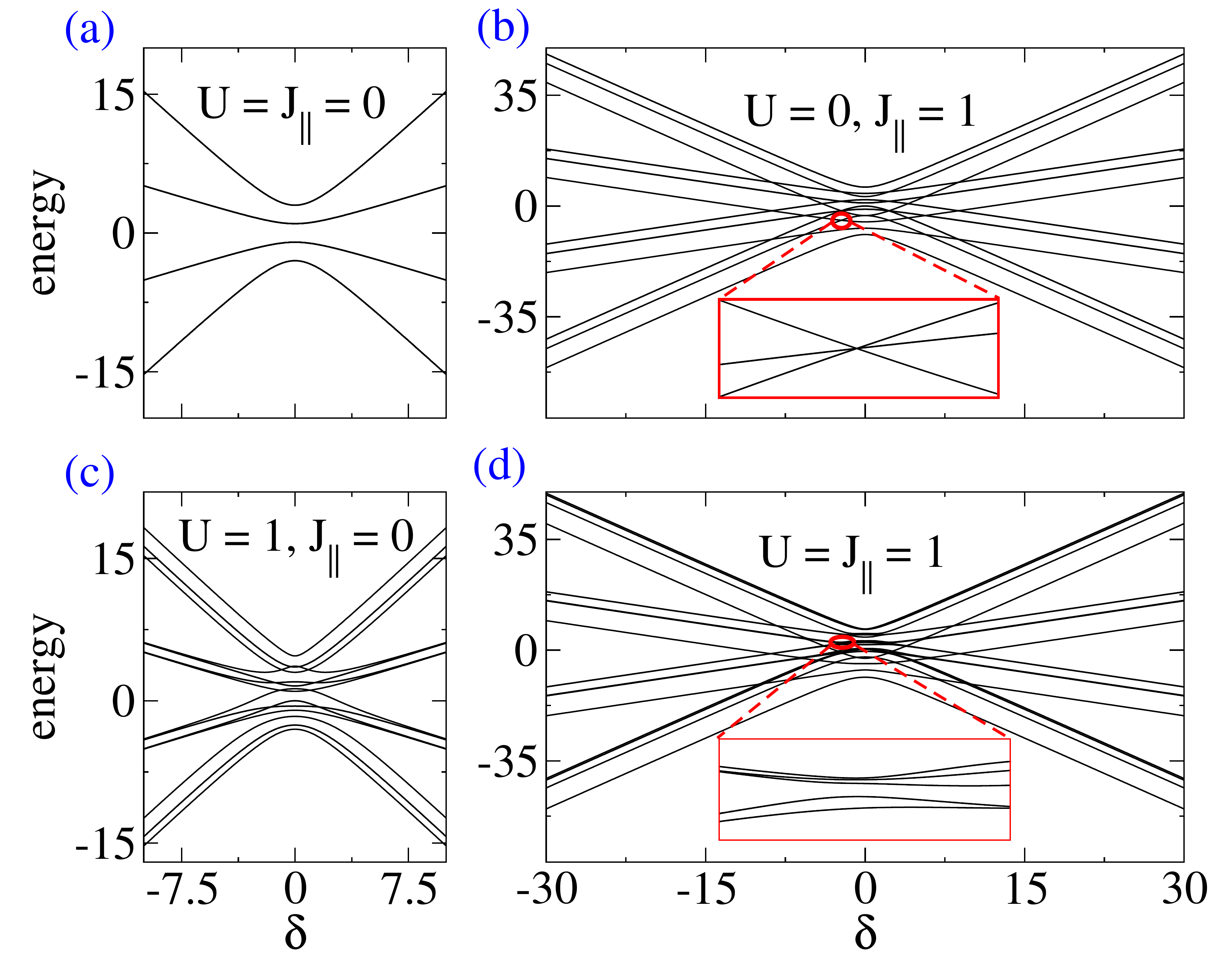}
\caption{  \label{fig:energy_spectra}
(color online) 
Energy spectra of Bose-Hubbard ladder for $L_s=N=3$ in zero momentum subspace.
(a) $U=\jp=0$: the eigenstates are degenerate.
(b) $\jp=1, U=0$: each band splits into eigenstates with different kinetic
energy.  Level crossings are real (not avoided) crossings, as shown in inset.
(c) $U=1, \jp=0$: the on-site interaction splits the eigenstates within each
band.
(d) $\jp=U=1$:  all crossings in the
spectrum are avoided, as shown in the inset.
}
\end{figure}

Figure \ref{fig:energy_spectra} shows energy spectra for $N=3$ bosons in a
$L_s=3$ (three-rung) ladder.  At large $\left|\delta_0\right|$, there are
$N+1$ groups of states (`bands'), corresponding to 0,1,2,...,$N$ bosons on the
left leg.  This is analogous to the dimer which has $N+1$ states.  Within each
of these bands, the states have splitting (dispersion) determined mainly by
the tunneling within the legs, $\jp$.  For $U=\jp=0$, the bands are each
degenerate (Figure \ref{fig:energy_spectra}a).

Figure \ref{fig:energy_spectra}b shows that in the $U=0$, $\jp\neq0$ case
there are true (unavoided) level crossings (Figure \ref{fig:energy_spectra}d
inset).  When both $U$ and $\jp$ nonzero, the crossings are all avoided
(Figure \ref{fig:energy_spectra}d inset).  In the sweeps considered in this
paper, going from $\pm\delta_0$ to $\mp\delta_0$, the system wavefunction
crosses a complex network of avoided level crossings such as that shown in
Figure \ref{fig:energy_spectra}d.

In Figure \ref{fig:n_R_as_function_of_time}a we show the paths truly adiabatic
sweeps would take, for both the g.s.\ sweep and the inverse sweep.  In the two
cases, the starting points are the lowest state of the lowest band (g.s.\
sweep) and the lowest state of the highest band (inverse sweeps).

\subsection{Sweeps and adiabaticity in few-rung ladders}

\begin{figure}
 \centering
\includegraphics[width=0.48\textwidth]{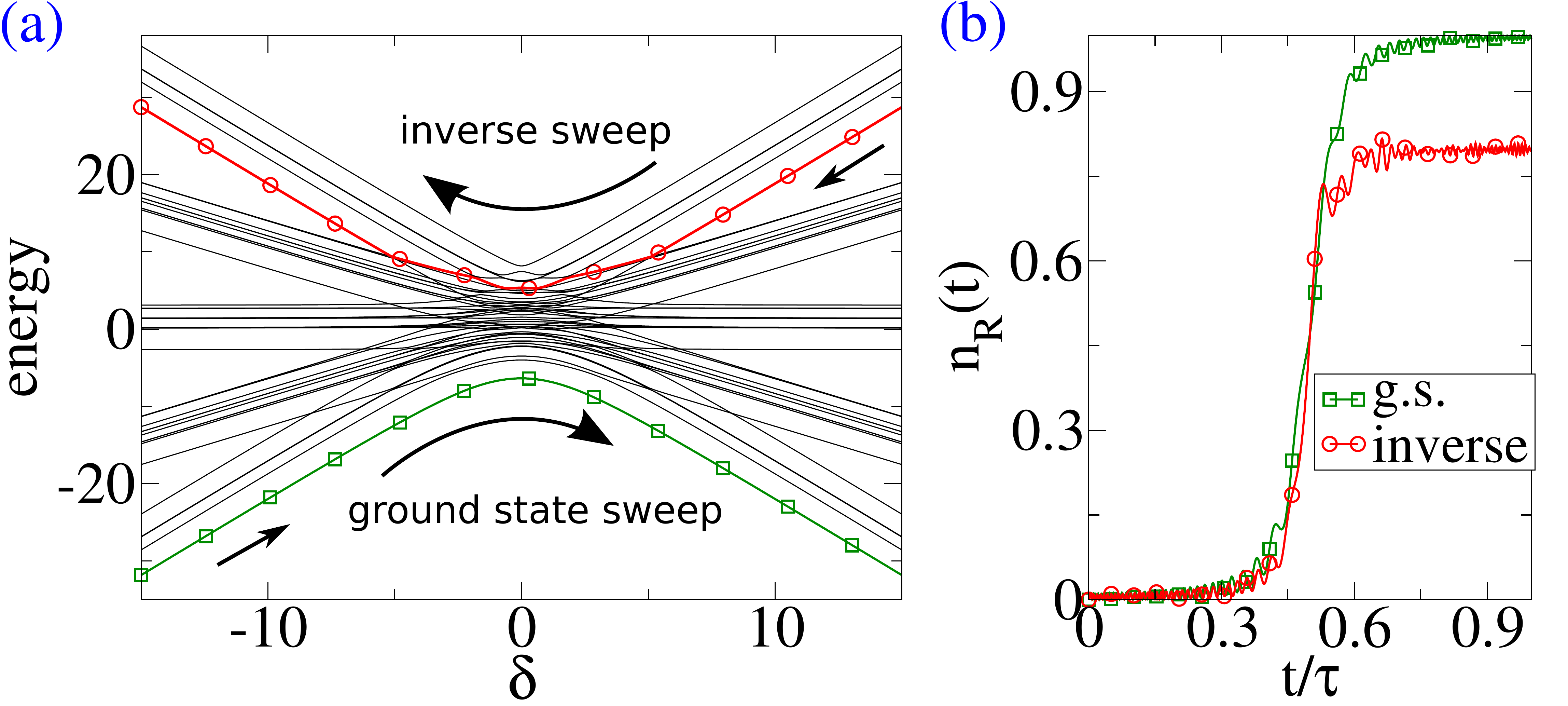}
\caption{ \label{fig:n_R_as_function_of_time}
(color online) 
(a) Energy spectrum in zero momentum subspace for $L_s=3$, $N=4$, $U=1$,
$\jp=0.4$, with 42 levels.  The adiabatic paths for g.s.\ and inverse sweeps
are highlighted.
(b) Fraction of bosons on the right leg $n_R(t)$ as function of time, during
sweeps with ($\delta_0$,$\tau$) = (20,40).  For the g.s.\ sweep the $n_R(t)$
reaches nearly unity, following the adiabatic path closely.  In contrast,
after the inverse sweep $n_R(t)$ falls far short of unity, even with a sweep
as slow as $\tau=40$.
}
\end{figure}

As we found for the dimer, in Bose-Hubbard ladders interactions are necessary
for seeing a difference in adiabaticity between ground state and inverse
sweep.  In the appendix \ref{sec:non_interacting}, we provide analytical
calculations for non-interacting ($U=0$) ladders where the g.s.\ and inverse
sweeps are identical and there is no breakdown phenomenon.

In this subsection and the rest of the main text, we will consider finite
interactions.  For $U\neq0$ ladders, the g.s.\ and inverse sweeps are
different (as for the dimer), and in addition there is also the breakdown
phenomenon.  In this subsection, we present the main features of the
non-equilibrium dynamics in few-rung Bose-Hubbard ladders during linear sweeps
of the energy bias, and also present the main features of the transfer
efficiency.

\begin{figure}[!t]
 \centering
\includegraphics[width=0.45\textwidth]{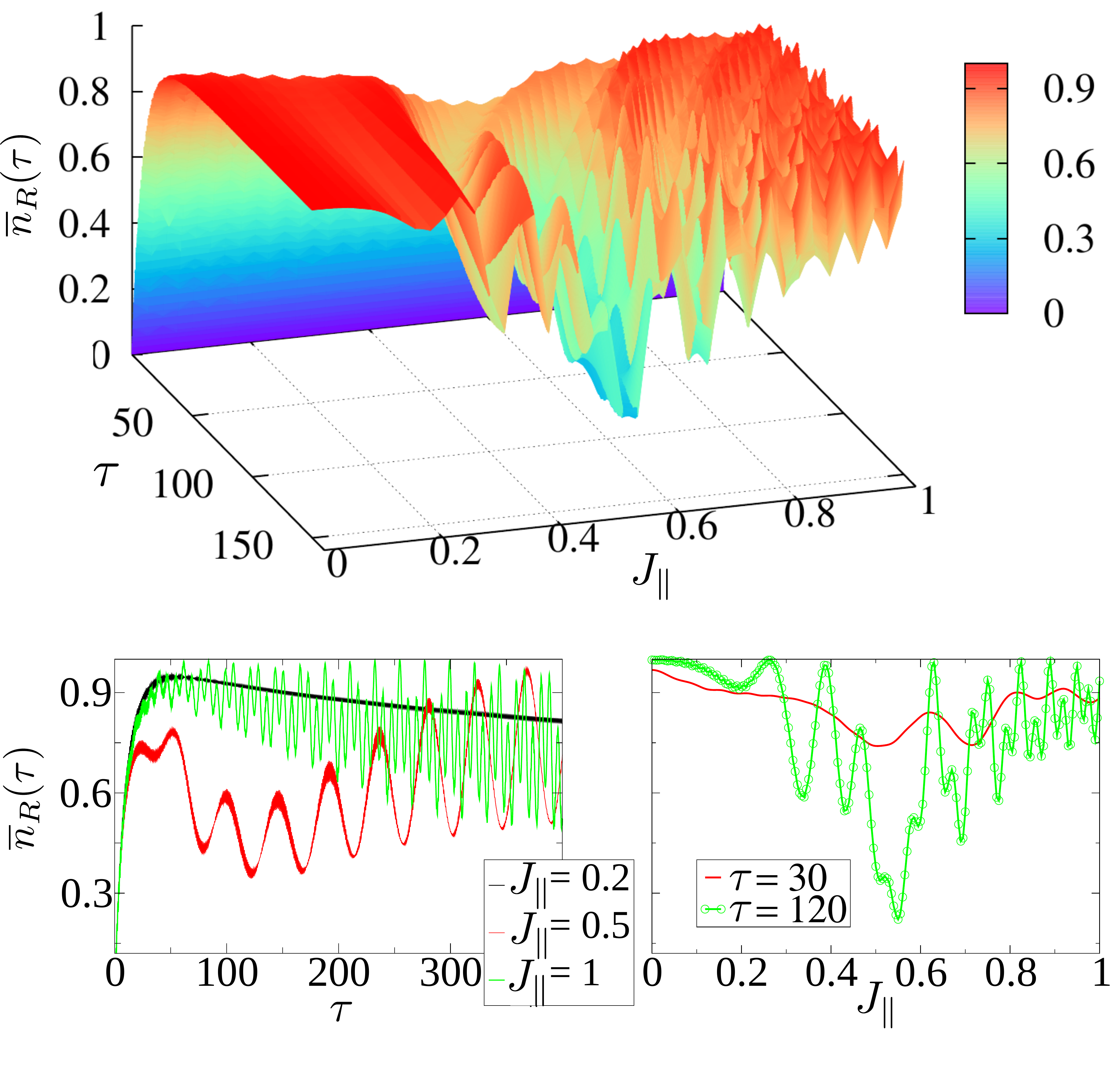}
\caption{  \label{fig:n_R_as_function_of_tau_L_2}
(color online) 
Transfer efficiency $\overline{n}_R(\tau)$ as function of the sweep time
$\tau$ and the intra-leg hopping $\jp$ for inverse sweep, for the ``minimal"
ladder $L_s=N=2$ and $U=1, \delta_0=30$.  Non-monotonicities with respect to
both $\tau$ and $\jp$ are visible in the 3D plot, and are also shown
separately in 2D panels.
}
\end{figure}

Figure \ref{fig:n_R_as_function_of_time}b shows the time dependence of the
boson fraction on the right leg $n_R(t)$, for a $\tau=40$ sweep.  For this
sweep time, the ground state sweep exhibits almost adiabatic behavior, while
the inverse sweep deviates significantly.  This difference is similar to that
described for the dimer (single-rung ladder) in section \ref{sec:dimer}.
Compared to the dimer, the fundamentally new feature in the ladder system is
seen in the dependence of $n_R(\tau)$ on the sweep time $\tau$, which is
non-monotonic and shows a pronounced region of decrease with increasing
$\tau$.  This is displayed in Figure \ref{fig:n_R_as_function_of_tau_L_2}.  

Oscillations in $n_R(t)$ complicate the behavior of the transfer efficiency
$n_R(\tau)$.  So, after the sweep, we evolve the final state with the time
independent final Hamiltonian $H(t=\tau)$ over a large additional time, to get
the time-averaged transfer efficiency $\nr:=\langle n_R(t>\tau)\rangle_t$.

Figure \ref{fig:n_R_as_function_of_tau_L_2} shows $\nr$ as a function of the sweep time $\tau$ and of the intra-leg tunnel
coupling $\jp$ for inverse sweeps in the smallest non-trivial ladder
($L_s=N=2$).  The transfer efficiency exhibits non-monotonicities as function
of both $\tau$ and $\jp$.  For faster sweeps $\nr$ increases with increasing
$\tau$.  In contrast, in the slow sweep regime ($\tau\gtrsim20$), $\nr$
decreases with increasing $\tau$ for $\jp\gtrsim{0.2}$.  In addition, for
$\tau\gtrsim20$, the transfer efficiency at fixed $\tau$ also depends
non-monotonically $\jp$.
These two non-monotonicities are analogs of those observed in experiment
(Figure \ref{fig:experiment_drawing}), and are analyzed in detail in the next
section.
Compared to the larger systems considered in Ref.~\cite{DMRG}, we have
additional oscillatory features, analyzed in section
\ref{sec:S_oscillations}.

\section{The non-monotonic behavior of the transfer efficiency} \label{sec:non-monotonicities}

In this section we provide microscopic explanations of the two
non-monotonicities of the transfer efficiency $\nr$ shown in
Figure \ref{fig:n_R_as_function_of_tau_L_2}.  The explanations are based on
considerations of the structure of energy spectra of Bose-Hubbard ladders and
the couplings between energy levels.  The $\tau$ non-monotonicity can be
understood by neglecting the crossing structure in the small $|\delta|$ regime
of the spectrum and only examining the coupling matrix elements between the
large-$|\delta|$ states.  This will be explained first
in \ref{subsec:explanation_nonmon_tau}. The $\jp$ non-monotonicity arises from
features of the crossing structure of the spectrum and is described
in \ref{subsec:explanation_nonmon_jp}.


\subsection{Non-monotonicity with sweep time} \label{subsec:explanation_nonmon_tau}

\begin{figure}
\centering
\includegraphics[width=0.98\columnwidth]{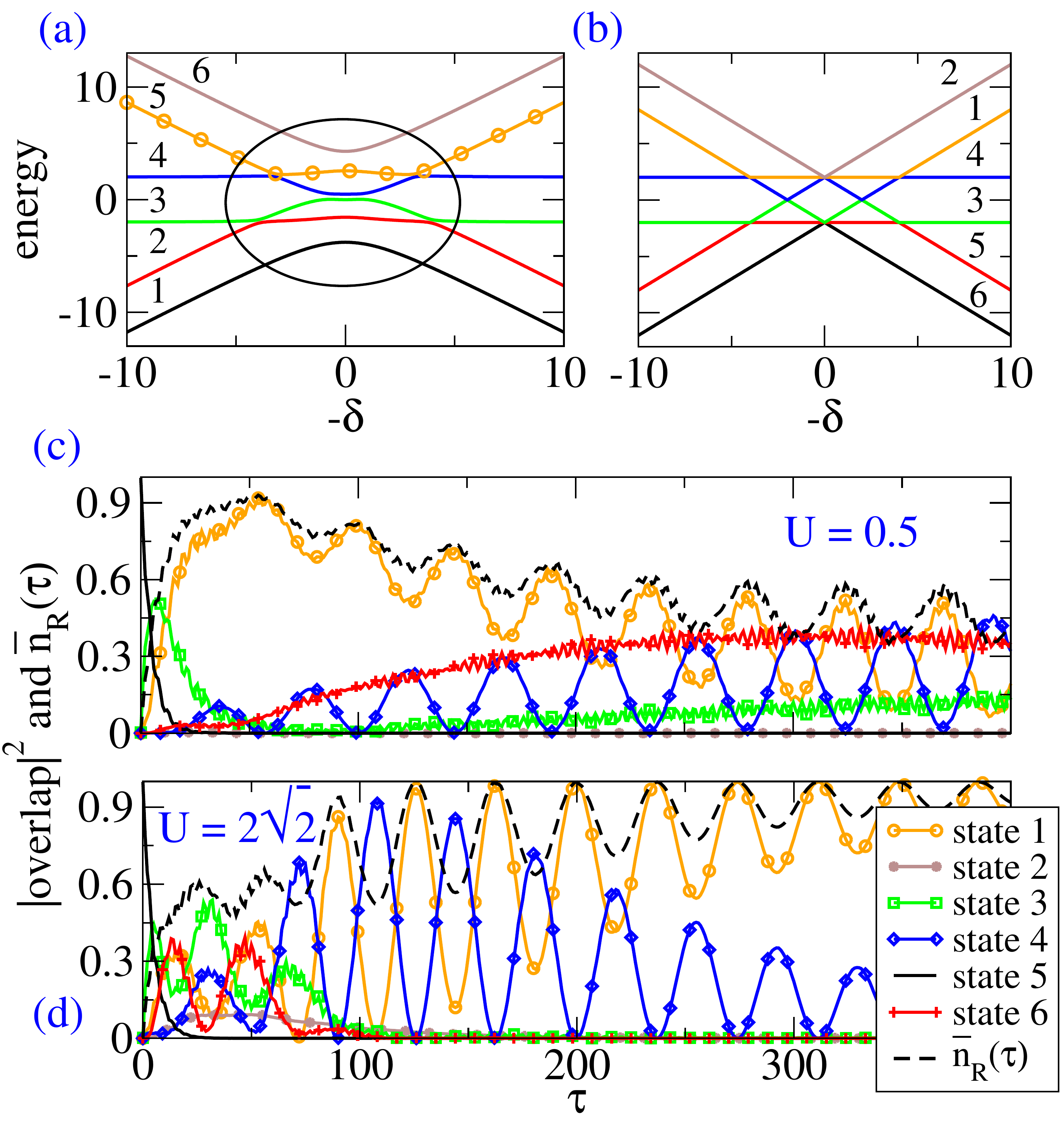}
\caption{ \label{fig:spectrum_L_N_2}
(color online) 
(a) Spectrum of minimal ladder ($L_s=N=2$), with $J_\parallel=0.5, U=1$.  The
adiabatic path for inverse sweep (5 $\rightarrow$ 1) is highlighted with
circles.  The crossing structure for small $|\delta|$ is indicated by an
ellipse.  Since $\delta(t=0) > 0$ we use $-\delta$ on the horizontal axis so
that the sweep goes from left to right.
(b) The diabatic basis states, obtained by setting $U=J_\perp=0$.
(c) Overlap of the final state with the eigenstates of the 
final Hamiltonian and transfer efficiency. Same parameters as in panel (a);
$\delta_0=30$. 
System goes into state 1 for fast sweeps, but population of state 1 is reduced
with increasing $\tau$ in favor of `band-excited' states 4 and 6.
(d) Larger interaction, for which there is no breakdown.  The final population
of state 1 increases monotonically (neglecting oscillations) with $\tau$.
}
\end{figure}

In this subsection we describe the microscopic origin of the non-monotonic
behavior of the $\nr$ function (breakdown of adiabaticity).  We will detail an
analysis for the simplest case of $L_s=N=2$ (two bosons in two-rung ladder,
Figure \ref{fig:spectrum_L_N_2}), and extract the general explanation from
this special case.  In terms of the state numbering in
Figure \ref{fig:spectrum_L_N_2}, adiabaticity in the inverse sweep involves
starting at state 5 on the left and populating state 1 on the right.
`Breakdown' involves a non-monotonic dependence on $\tau$ of the final state 1
population.  We therefore examine and explain the overlaps of the final
wavefunction with the final eigenstates, as a function of $\tau$.

The details of the crossing structure of the spectrum at small $|\delta|$
(ellipse in Figure \ref{fig:spectrum_L_N_2}a) vary widely with Hamiltonian
parameters (c.f.\ Figure \ref{fig:energy_spectra}).  On the other hand, the
breakdown effect is robust through a large region of parameter space.  This
suggests that the phenomenon can be explained ignoring the exact crossing
structure of the spectrum around $\delta=0$.  Our strategy is therefore to use
the states at large $\delta$ as a `diabatic' basis and examine the coupling
matrix among these states.
%



\textbf{Diabatic representation:}
The spectrum (Figure \ref{fig:spectrum_L_N_2}a) consists of three bands of two
states each, and the initial state is very close to the lowest state of the
highest band, i.e. state 5.

The time evolution starts and ends at $|\delta(t=\{0,\tau\})| \gg 1$, where
the Hamiltonian \eqref{eq:Hamiltonian_ladder} is dominated by the energy bias
term $\delta/2 \sum_i (\hat{n}_{i,L}-\hat{n}_{i,R})$.  If one regards the
on-site interaction ($U$) and the inter-leg hopping ($J_{\perp}$) terms as
couplings, then in analogy to the two-level Landau-Zener problem the
eigenstates of the Hamiltonian with $U=J_\perp=0$ can be regarded as the
diabatic basis.  The spectrum of the coupling free Hamiltonian
$H_{J_\perp=U=0}$ is shown in Figure \ref{fig:spectrum_L_N_2}b.  The
diabatic basis can be calculated analytically using the procedure described in
Appendix \ref{sec:non_interacting} for treating non-interacting Hamiltonians.

For large energy bias $|\delta| \gg 1$ the eigenstates of the full Hamiltonian
coincide with the diabatic states.  The coupling strengths between diabatic
states sets the probability for transition between these states. 
Loosely speaking, this means the overlap of the final wavefunction with a
diabatic state after the sweep is larger when the coupling (either direct or
higher-order) with the initial state is larger.  In other words, a transition
with larger coupling should be noticeable already for faster sweeps (smaller
$\tau$), while states weakly coupled would need a slower sweep to get
populated.  
%
 
The coupling matrix between the diabatic states is obtained by
representing the full Hamiltonian in the diabatic basis:
%
%
%
\begin{equation}\nonumber
\begin{pmatrix}
- & \frac{U}{2} & -\sqrt{2} J_\perp & 0 & 0 & 0 \\
 \frac{U}{2} & - & 0 & -\sqrt{2} J_\perp & 0 & 0 \\
 -\sqrt{2} J_\perp & 0 & - & 0 & -\sqrt{2} J_\perp & 0 \\
 0 & -\sqrt{2} J_\perp & 0 & - & 0 & -\sqrt{2} J_\perp \\
 0 & 0 & -\sqrt{2} J_\perp & 0 & - & \frac{U}{2} \\
 0 & 0 & 0 & -\sqrt{2} J_\perp & \frac{U}{2} & - \\
\end{pmatrix}.
\label{eq:Hamiltonian_matrix_adiabatic_basis}
\end{equation}   
%
%
%
The diabatic basis states are ordered according to the numbering of
Figure \ref{fig:spectrum_L_N_2}.  
We note that states from neighboring bands with same kinetic energy along the
ladder (i.e., states $1;3;5$ and states $2;4;6$) are coupled by $J_\perp$.
States from the same band, if coupled, are coupled through the interaction
$U$.  We will next show how these observations on coupling terms allow an
explanation of the breakdown phenomenon, and also allow us to predict regions
of parameter space where the phenomenon does not occur.



\begin{figure}
\centering
\includegraphics[width=0.48\textwidth]{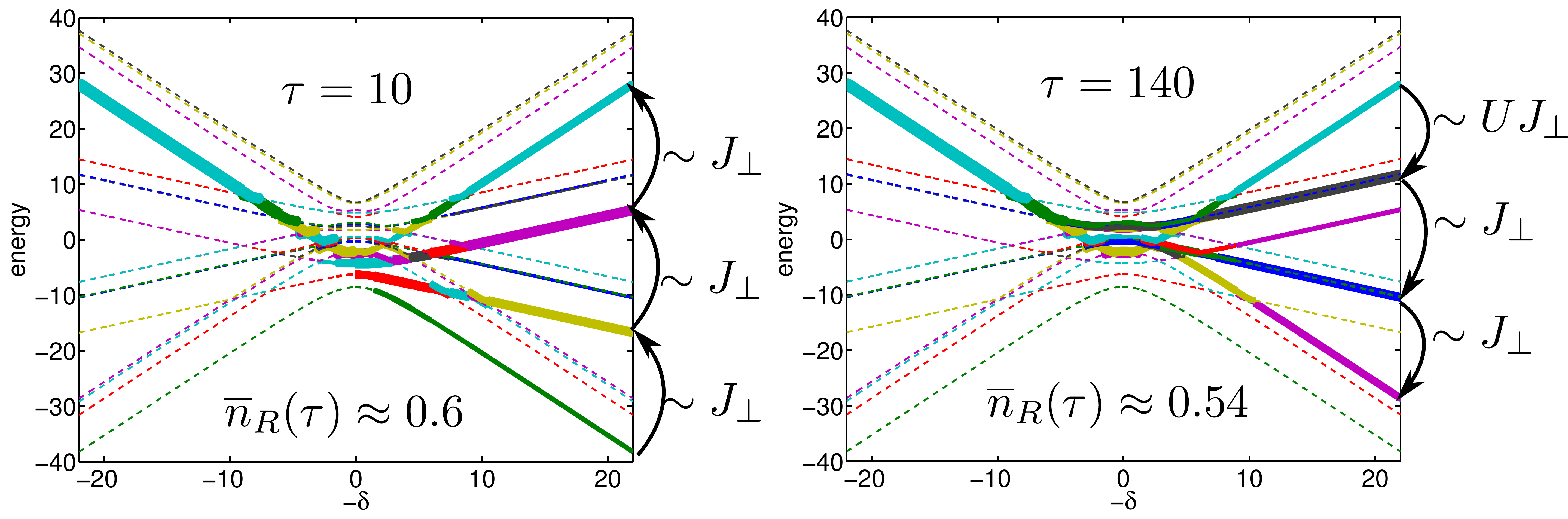}
\caption{ \label{fig:spectrum_evolution}
(color online) 
Spectra for $L_s=N=3$, $\jp =U=1$, showing time evolution during inverse
sweeps ($\delta_0=30$) through overlaps, which are indicated by line widths.
The width of lines is $\sim \ln(\text{overlap})$, and if the overlap is
smaller than 2\% the line is dashed.  Transfer efficiency has a maximum at
$\tau=40$, $\nr=0.87$.  As argued in text, we see that for small $\tau$ (left)
the main process is the successive excitation of band ground states.  In the
breakdown regime (right), the higher states of lower bands are excited.
Arrows highlight this mechanism pictorially.
} 
\end{figure}

\textbf{Breakdown of adiabaticity:}
In Figure \ref{fig:spectrum_L_N_2}c we show the final overlaps as a function
of $\tau$, for a parameter combination for which breakdown is exhibited.  At
fast ramp rates (small $\tau$), the states at the bottom of bands (states 1,3)
get occupied, so that the occupancy of state 1 (and hence $\nr$) increases to
a maximum.  For slower sweeps, the occupancy of these states is lost to
higher-momentum states in lower bands, (states 4,6).  This decrease of state 1
occupancy (and hence of $\nr$) is the breakdown phenomenon.  The same
description holds for larger ladders (which have more bands and more states in
each band) --- (a) the band ground states are successively occupied at small
times, so that the $\nr$ and the lowest state of the highest band increase
initially with $\tau$, and (b) for slower ramps excited states of lower bands
gain weight at the expense of the highest band g.s.  These processes are
illustrated for a larger system in Figure \ref{fig:spectrum_evolution}.

We can explain this in terms of two different coupling strengths (different
time scales) for initial excitation of the highest band ground state (state 1)
and for later transfer to excited states of lower bands.  In the $L_s=N=2$
case, this relies on $\frac{1}{2}U$ being smaller than $\sqrt{2}J_{\perp}$,
and for larger ladders we would require $U$ to be smaller than $c_1J_{\perp}$
for some constant $c_1$.  The lowest states of successive bands can get
populated with coupling strength $J_{\perp}$, so that the time scale for
initially populating the `adiabatic' final state (state 1) is set by
${\sim}J_{\perp}^N$.  The time scale for moving to excited states of lower
bands is however set by the weaker coupling ${\sim}J_\perp^{N-1}U$, and is
thus slower, and can happen only for slower sweeps.  This explains the loss of
state 1 occupancy in favor of excited states of lower bands, at larger $\tau$.

Population of band excited states is fundamental to the breakdown mechanism.
This corresponds to the idea of longitudinal low-lying excitations (gapless
phonon excitations) along legs, highlighted in Ref.~\cite{experiment}.  Note
that our microscopic mechanism above does not rely on any idea of gapless
excitations, which strictly speaking only occurs in infinitely long ladders.

\textbf{Parameter regimes without breakdown of adiabaticity:}
A strong confirmation of our picture is the absence of breakdown at larger
$U$, as shown in Figure \ref{fig:spectrum_L_N_2}d.  For larger $U$, the time
scale for populating lower-band excited states is comparable or smaller than
the time scale for initial population of higher-band ground state.  Not having
a slower time scale, there is no longer any mechanism for $\nr$ decrease with
$\tau$.

Another regime with no breakdown is that of small $\jp$.  A simple way to
understand this is that, for $\jp\rightarrow0$, the system corresponds to
isolated single dimers, and we have seen in section \ref{sec:dimer} that the
phenomenon is an absent in a single dimer.


\subsection{Non-monotonicity with intra-leg hopping}   \label{subsec:explanation_nonmon_jp}

\begin{figure}
\centering
\includegraphics[width=0.96\columnwidth]{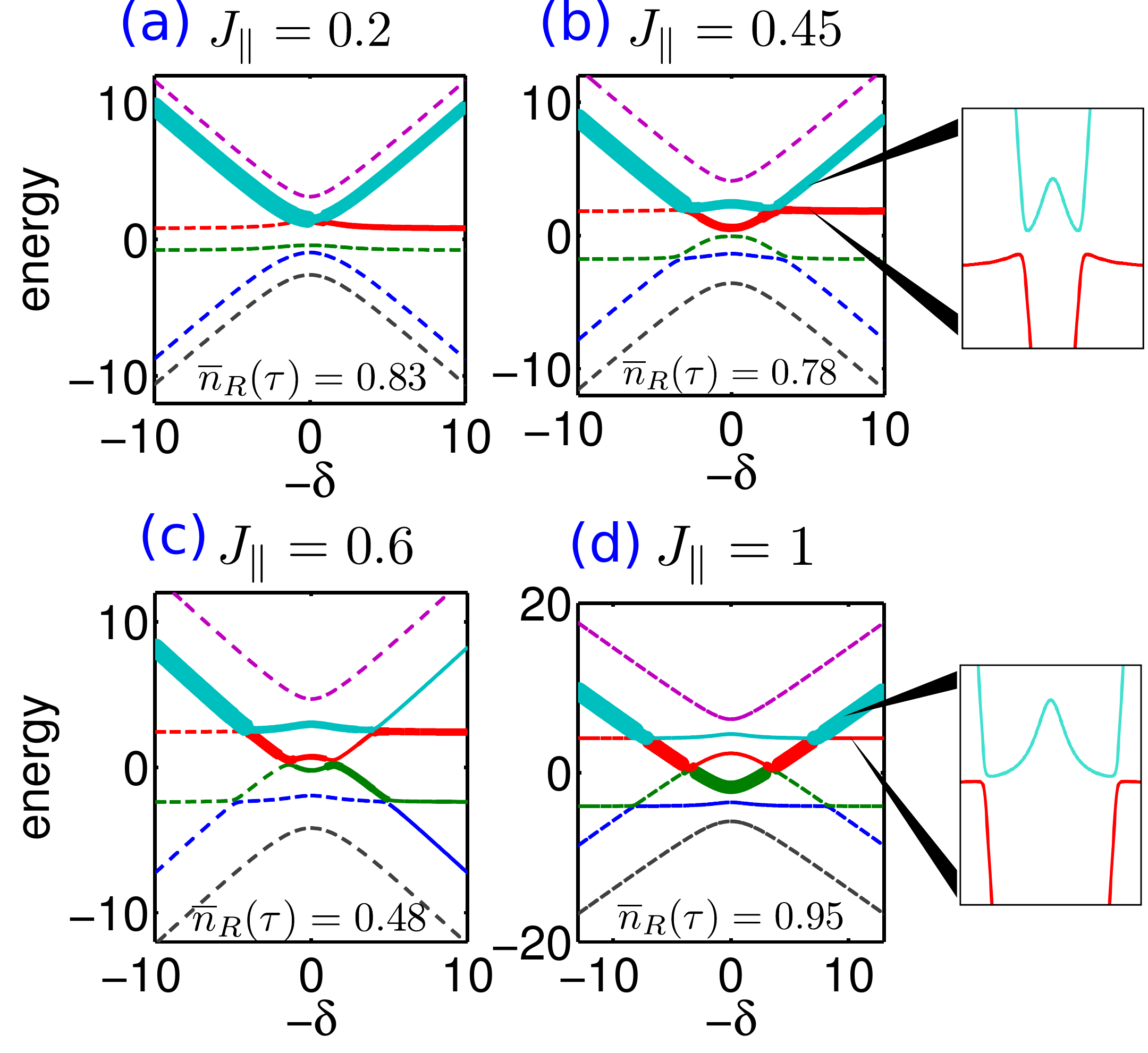}
\caption{  \label{fig:spectrum_evolution_5}
(color online)
Time evolution during inverse sweep in the minimal ladder ($L_s=N=2$), shown
through overlaps indicated by line widths.  The width of lines is $\propto$
overlap, and if the overlap is smaller than 5\% the line is dashed.
Here $U=1$, $\tau = 200$, $\delta_0=20$.
Insets on right show blowups of the crossings in energy spectra, in the same
scale, highlighting the differences between small and large $\jp$.  
}
\end{figure}

To explain the non-monotonicity of $\nr$ with $\jp$ at moderately large
$\tau$, we now consider the crossing structure of the energy spectrum at small
$|\delta|$ and how this structure changes with $\jp$.  
Figure \ref{fig:spectrum_evolution_5} shows this for the minimal ($L_s=N=2$)
ladder.

For very small $\jp$ (Figure \ref{fig:spectrum_evolution_5}a) the bands have
little or no overlap, so the lowest state of the highest band has little or no
proximity to any other energy level during the sweep.  As a result, the
transfer efficiency is large for moderate $\tau$.

As $\jp$ is increased, the width of each band increases, resulting in multiple
crossings, in particular between the lowest state of the highest band and
various excited states of lower bands (Figures \ref{fig:spectrum_evolution_5}b
and \ref{fig:spectrum_evolution_5}c).  Via these crossings, the wavefunction
can populate states of lower bands, leading to loss of transfer efficiency.

When $\jp$ is increased even further (Figure \ref{fig:spectrum_evolution_5}d),
the crossings happen at larger $|\delta|$ and have smaller gaps.  (Physically,
the gaps are smaller at larger $|\delta|$ because the diabatic states are more
sharply different and hence mix less.)  Small-gap crossings behave more like
true crossings.  Thus at large $\jp$ and moderate $\tau$ the main path taken
by the wavefunction ends up at the lowest level of the highest band, even
though this path passes through many crossings, as shown in
Figure \ref{fig:spectrum_evolution_5}d.

To summarize, increasing $\jp$ has two effects: at first it increases the
number of crossings, which reduces the transfer efficiency, and then for even
larger $\jp$ it makes the crossing gaps smaller, which again increases the
transfer efficiency.  This explains the overall minimum in the $\nr$ versus
$\jp$ behavior.  Of course, the behavior is complicated by interference
effects (St\"uckelberg oscillations), prominent in the lower right panel of
Figure \ref{fig:n_R_as_function_of_tau_L_2}.  In
Figure \ref{fig:spectrum_evolution_5} we have chosen four $\jp$ values that
highlight the overall trend of the $\nr$ versus $\jp$ behavior.


\section{St\"uckelberg oscillations} \label{sec:S_oscillations} 

\begin{figure}
\centering
\includegraphics[width=0.48\textwidth]{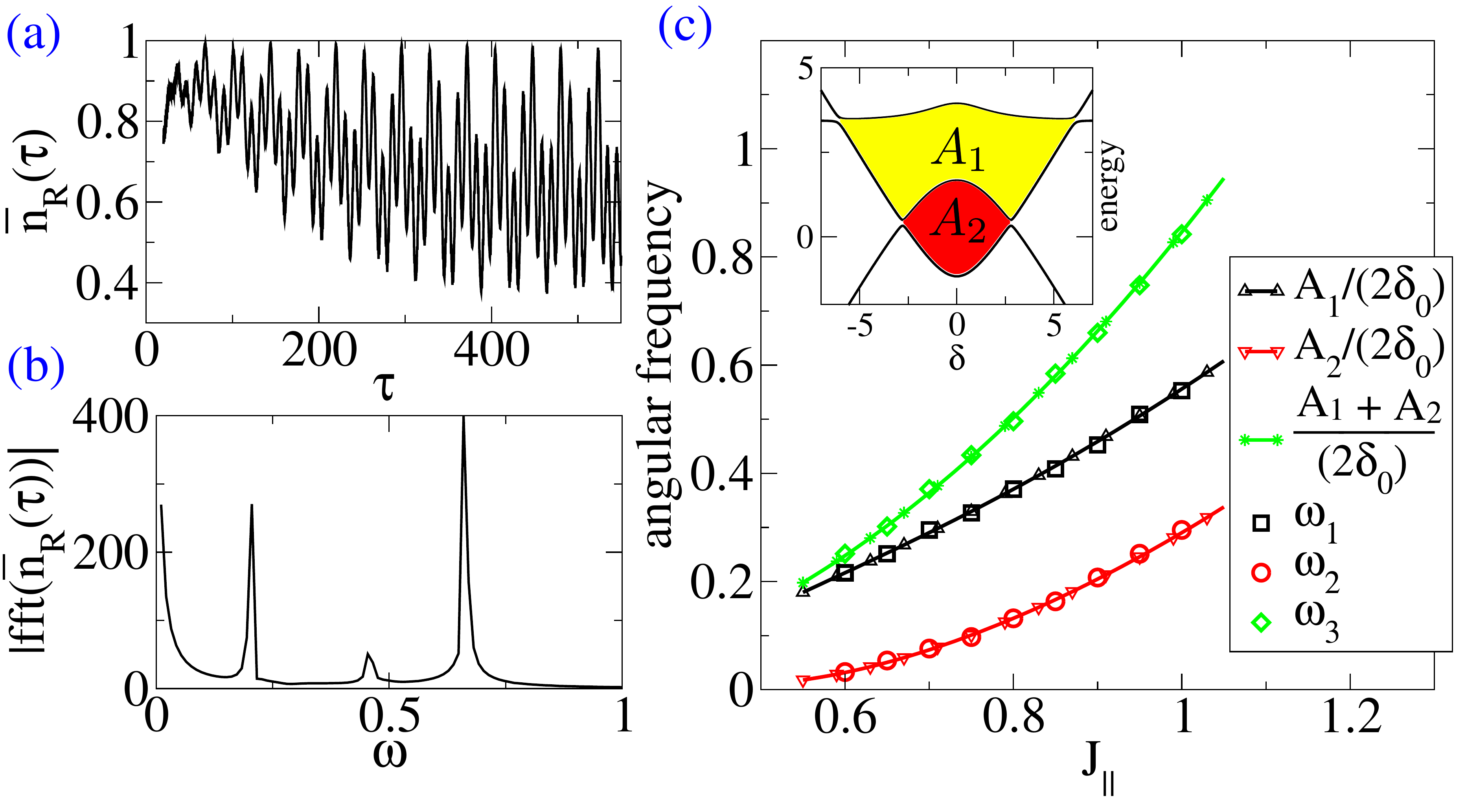}
\caption{  \label{fig:stukelberg_fourier}
(color online) 
St\"uckelberg oscillations in minimal ladder ($L_s=N=2$), with $\delta_0=20$,
$U=1$, $\jp = 0.85$.
(a,b) Transfer efficiency $\nr$ and its Fourier transform.
(c) Frequencies compared to areas enclosed by crossing lines in the spectrum.
Inset shows relevant region of spectrum and definitions of areas. 
}
\end{figure}

In this section we analyze the oscillations observed in the transfer
efficiency
(Figures \ref{fig:n_R_as_function_of_tau_L_2}, \ref{fig:spectrum_L_N_2},
and \ref{fig:stukelberg_fourier}a).  While the overall trends discussed
previously reflect the macroscopic physics which is the focus of
Refs.~\cite{experiment,DMRG}, the oscillations are particular to our
microscopic (small-system) ladders.

When a system is swept through a parameter region containing multiple avoided
level crossings, the wavefunction may split into two eigenstates at one
crossing and, later in the sweep, levels with nonzero weight may meet again
and interfere.  Such interference can lead to oscillatory behavior of
observables as a function of the sweep rate \cite{Stuckelberg_interferometry}.
The wavefunction component following the energy level $E_i$ will accumulate
the phase $\int{E_i}dt$.  As a result, in quantities like $\nr$ there are
interference terms carrying phase factors of type
$\int\left[E_i-E_j\right]dt$, for each pair of levels ($i$,$j$) with nonzero
weight.  Since the sweeps change $\delta$ linearly in time, $dt =
-\frac{\tau}{2 \delta_0} \text{d} \delta$.  Thus the phase factors are
\begin{equation}
\frac{\tau}{2\delta_0} \int \left[ E_{i}(\delta)-E_{j}(\delta)\right]\text{d} \delta = \frac{A_{\alpha}}{2\delta_0} \tau
\end{equation}
where $A_{\alpha}$'s are the areas enclosed between levels and crossings in
the energy-$\delta$ plane, as shown in Figure \ref{fig:stukelberg_fourier}c
inset.  Thus the oscillations as a function of $\tau$ carry frequency
$A_{\alpha}/2\delta_0$.

In Figure \ref{fig:stukelberg_fourier}b the Fourier spectrum of $\nr$ is
shown.  There are three pronounced peaks.  These frequencies can be compared
to ``enclosed areas'' in the energy spectrum diagram
(Figure \ref{fig:stukelberg_fourier}c inset).
Figure \ref{fig:stukelberg_fourier}c shows excellent agreement between the
oscillation frequencies, obtained by Fourier analysis, and the exact
numerically calculated areas $A_{\alpha}$ multiplied by $\frac{1}{2\delta_0}$.

\section{Summary and Conclusion}  \label{sec:summary_conclusion}

A recent experiment with trapped bosonic atoms in ladder-shaped optical
lattices has highlighted the intricacies that can arise in the approach to
adiabaticity in interacting systems \cite{experiment, DMRG}.  Motivated by
this experiment, we have undertaken an exact study of Bose-Hubbard ladders of
explicitly finite size.  We have thoroughly examined the energy spectra, their
crossings, and the transitions between them, and explained some non-trivial
features of the transfer efficiency.

The understanding emerging from this analysis is that, while interaction
effects are vital for the ``breakdown of adiabaticity'' and related effects,
the thermodynamic limit is not at all necessary for understanding the main
effects qualitatively.  In particular, the two non-trivial non-monotonicities
are both well-explained without appealing to concepts like gapless-ness of
excitation spectra or nonlinearities in mean-field descriptions.  

The present analysis opens up various avenues of possible future research.
Since a trapped system contains ladders of all possible sizes, one might ask
whether from the data one can extract information relevant to the sizes we
have considered here, in particular whether it is possible to study the
interference features (St\"uckelberg oscillations) which are prominent in
small ladders.  Second, non-equilibrium dynamics in the size regime between
the present sizes and large sizes remains unexplored.  This relates to the
widely appreciated difficulty of interpolating between full quantum treatments
of small systems and mean-field treatments of larger systems.

\appendix


\section{Non-interacting Bose-Hubbard ladder}\label{sec:non_interacting}

With periodic boundary conditions (translation invariance), the time
dependence of the $U=0$ Bose-Hubbard ladder of arbitrary length can be reduced
to a single 2x2 system, i.e., a single Landau-Zener type problem.

Writing the Hamiltonian \eqref{eq:Hamiltonian_ladder} with $U=0$ in momentum
space, one can diagonalize using a unitary transformation for every momentum
mode ($q = 0, \ldots , L_s-1$):
\begin{equation}
 \begin{split}
&\alpha_q = u_q \cdot a_{q,L} + v_q \cdot a_{q,R}  \\
&\beta_q =  v_q \cdot a_{q,L} - u_q\cdot a_{q,R} ,
 \end{split}
\end{equation}
with $u^2+v^2=1$.  The diagonalization condition yields $u/v =
2J_{\perp}/\delta$, and the resulting diagonal Hamiltonian is   
\begin{equation}
\begin{split}
 H^{U=0} =& \sum_q \bigg[ -2J_\parallel \cos(\frac{2 \pi}{L_s}q) (\alpha^\dagger_q\alpha_q + \beta_q^\dagger\beta_q) \\
		    & + \frac{\delta}{|\delta|} \frac{1}{2} \sqrt{\delta^2 + 4J_\perp^2} ( \beta_q^\dagger\beta_q - \alpha^\dagger_q\alpha_q)\bigg].
\end{split}
\label{eq:Hamiltonian_after_transformation} 
\end{equation}
The eigenstates can now be created by applying $\alpha_q^{\dagger}$,
$\beta_q^{\dagger}$ operators on the vacuum, e.g., the ground state at
negative $\delta$ is   $|\psi_{g.s.}\rangle
=\frac{1}{\sqrt{N!}}(\beta_0^\dagger)^N|0\rangle$.  Dynamics is obtained using
Heisenberg equations of motion for the $\beta_q$ operator, which leads to the
same 2x2 problem for each $q$ mode:
\begin{equation}
i \hbar \begin{pmatrix} \dot{v}\\\dot{u}\end{pmatrix} = \begin{pmatrix}\frac{\delta(t)}{2}-2J_\parallel & -J_\perp\\ -J_\perp & -\frac{\delta(t)}{2}-2J_\parallel  \end{pmatrix} \begin{pmatrix} v\\u \end{pmatrix} 
\label{eq:diff_eq_coefficients_ladder}                                                    
\end{equation}
The fraction of bosons in the right leg is $n_R(t) = |u(t)|^2$.  The 2x2
dynamics is too simple for $n_R(t)$ to display any of the nontrivial behaviors
we have studied in the main text with nonzero interactions.  For large
$\delta_0$ and $\tau$, we can use the Landau-Zener formula to calculate the
transfer efficiency: $n_R(\tau) = 1 - \exp[-2\pi{J_\perp^2}/\alpha]$, with
$\alpha = \pm 2 \frac{\delta_0}{\tau}$.  Of course, this increases
monotonically with $\tau$.

\begin{figure}[!b]
\centering
\includegraphics[width=0.98\columnwidth]{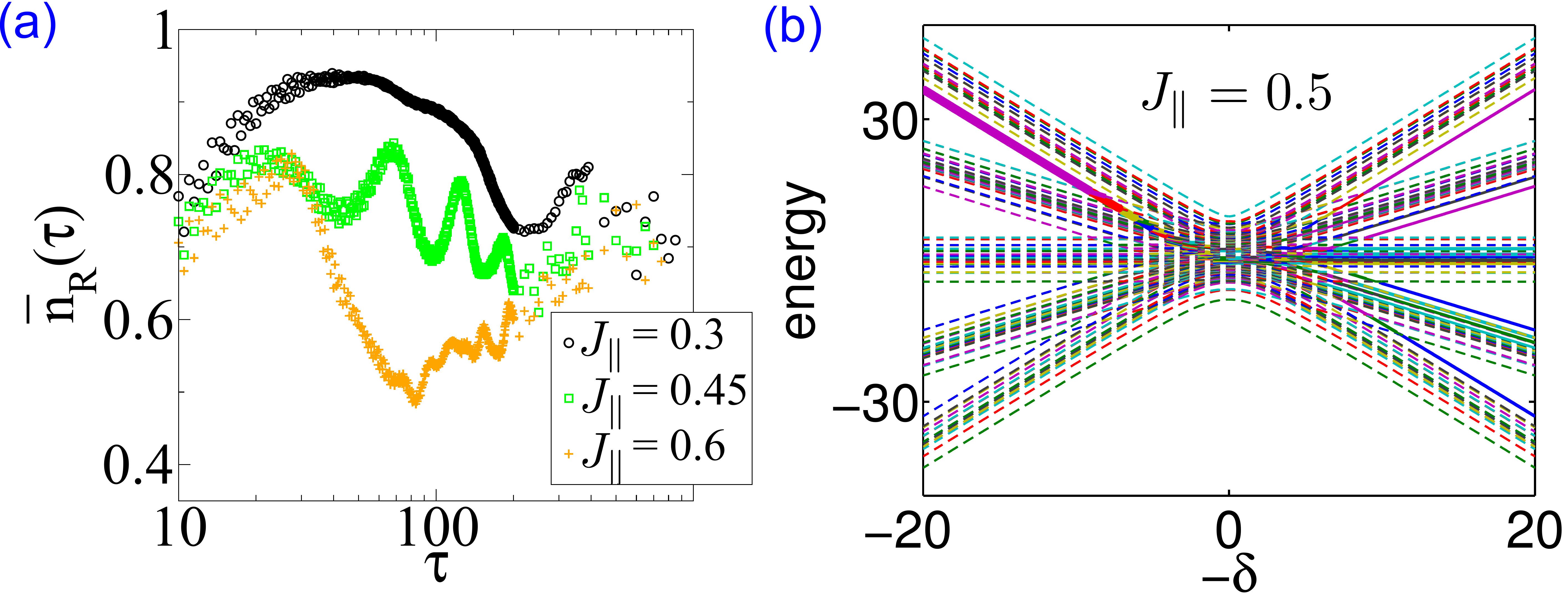}
\caption{ \label{fig:n_R_L_5}
(color online) 
(a) Breakdown of adiabaticity in larger system. $L_s=5$, $N=4$, 
$U=1$, $\delta_0=20$.  
(b) Time evolution during an inverse sweep in the breakdown regime ($\tau =
80$. $\jp = 0.5$), presented (as in Figures \ref{fig:spectrum_evolution}
and \ref{fig:spectrum_evolution_5}) through the line-widths representing
overlap ($\sim \ln(\text{overlap})$).
}
\end{figure}


\section{$L_s=5, N=4$}\label{sec:L_s_5}

In Figure \ref{fig:n_R_L_5} the breakdown of adiabaticity is shown for a
larger system containing four bosons in five-dimer ladder.  The St\"uckelberg
oscillations in $\nr$ are now far less pronounced, but otherwise the features
of the breakdown phenomenon are very similar to those presented in detail for
smaller systems in the main text.  The overlaps of Figure \ref{fig:n_R_L_5}b
show that band excited states, corresponding to longitudinal modes along the
leg directions, are excited in the breakdown process, consistent with the
picture developed in section \ref{subsec:explanation_nonmon_tau}.


\end{document}